# Semiquantum proxy blind signature based on quantum teleportation

Xiao Tan, Tian-Yu Ye*

College of Information & Electronic Engineering, Zhejiang Gongshang University, Hangzhou 310018, P.R.China

E-mail：yetianyu@zjgsu.edu.cn

**Abstract:** In this paper, we propose a novel semiquantum proxy blind signature scheme with quantum teleportation based on $\chi$ states, where the original message owner, the proxy signer and the third party are quantum participants with complete quantum capabilities, while the original signer and the signature verifier are semiquantum participants with limited quantum capabilities. It turns out that our protocol not only has complete blindness, unforgeability, non-repudiation and but also can resist the attack behavior from an eavesdropper. Compared with many previous quantum proxy blind signature protocols, our protocol may need less quantum resources and be easier to implement in reality, since both the original signer and the signature verifier are semiquantum participants with limited quantum capabilities.

## 1  Introduction

It is well known that digital signature can ensure the authenticity, the integrity and the non-repudiation of message transmission [1]. In a digital signature scheme, the legal signer uses her private key to generate a valid signature, and others uses their public keys to verify this signature. With the development of powerful quantum parallel computing, all seemingly complex classical digital signature schemes may be not safe anymore.

Quantum cryptography, put forward by Bennett and Brassard [2] in 1984, can be regarded as the combination of classical cryptography and quantum mechanics, and can provide the unconditional security in theory. After quantum mechanics was absorbed into classical digital signature, quantum digital signature was derived and hereafter has been developed into different branches with different functions, such as quantum arbitrated signature (QAS) [3-8], quantum blind signature (QBS) [9,10], quantum proxy signature [11][11], quantum proxy blind signature [12-18], etc. Apparently, quantum proxy blind signature is a special type of quantum digital signature. In a quantum proxy blind signature scheme, the message owner doesn't want the original signer to get the real content of her message, but only wants the original signer to give signature; and the original signer always cannot sign the message directly, and has to distribute the authority of signing task to one or more proxy signers. In 2014, Cao et al. [12] constructed a quantum proxy weak blind signature scheme. In 2017, Yan et al. [13]

designed a quantum multi-proxy weak blind signature protocol by using entanglement swapping. In 2019, Liang et al. [14] proposed a quantum multi-proxy blind signature scheme with four-qubit cluster states; Liu et al. [15] proposed a quantum group proxy blind signature scheme by using five-qubit entangled states; and Zhang et al.[16] [16] put forward a quantum proxy blind signature scheme which adopts a random sequence to blind the message. In 2020, Niu et al. [17] constructed a quantum proxy blind signature scheme with superdense coding. In 2022, Chen et al. [18] proposed a novel quantum multi-proxy blind signature scheme by using cluster states.

In a quantum cryptography protocol, all participants are always required to possess complete quantum capabilities. Unfortunately, it may be impractical, as not all of them have the ability to afford expensive quantum resources and operations in certain circumstances. In order to solve this problem, Boyer et al. [19] proposed the brand-new concept of semiquantumness in 2007. According to Refs.[19,20], in a semiquantum cryptography protocol, the party who has the abilities to perform all quantum operations is called quantum party, and the semiquantum party can only perform limited quantum operations, such as measuring qubits with the $Z$ basis (i.e., $\{|0\rangle,|1\rangle\}$), preparing qubits within the $Z$ basis, transmitting qubits and rearranging the order of qubits. Soon after the concept of semiquantumness was derived, by introducing it into quantum digital signature, scholars quickly established a special branch of semiquantum cryptography called as semiquantum digital signature. In 2019, Zhao et al.[21] [21] proposed a semiquantum bi-signature protocol based on W states, in which two signers sign a message at the same time by using quantum teleportation based on W states. In 2020, Zheng et al. [22] proposed a semiquantum proxy signature scheme by using teleportation based on quantum walk. In 2021, Xia et al.[23] [23] proposed a semiquantum blind signature protocol with five-particle GHZ states. In 2022, Wang et al. [24] put forward a semiquantum blind signature protocol immune to collective noise.

In this paper, we are devoted to designing a novel semiquantum proxy blind signature protocol with quantum teleportation based on $\chi$ states. Our protocol is proven to possess complete blindness, unforgeability, unforgeability and security. Compared with many previous quantum proxy blind signature protocols, our protocol may need less quantum resources and be easier to implement in reality, since both the original signer and the signature verifier are semiquantum participants with limited quantum capabilities.

## 2  Protocol description

In this section, we first construct a novel quantum teleportation scheme based on $\chi$ state, and then put forward a novel semiquantum proxy blind signature protocol with quantum teleportation based on $\chi$ states.

## 2.1 Quantum teleportation based on $\chi$ state

The $\chi$ states are four-particle maximally entangled states. In this paper, the following $\chi$ state [25] is used as the quantum carrier:

$$|\chi^{00}\rangle_{1234} = \frac{1}{2\sqrt{2}}(|0000\rangle+|0011\rangle-|0101\rangle+|0110\rangle+|1001\rangle+|1010\rangle+|1100\rangle-|1111\rangle)_{1234}. \quad (1)$$

Suppose that at the Sender's site, the unknown arbitrary single-particle state to be transmitted is

$$|\xi\rangle_m = a|0\rangle_m + b|1\rangle_m. \quad (2)$$

Here, both $a$ and $b$ are complex numbers which satisfy

$$|a|^2 + |b|^2 = 1. \quad (3)$$

Moreover, suppose that the Sender, Assistant 1, Assistant 2 and the Receiver share the four-particle entangled state shown in Eq.(1). Concretely speaking, the Sender has particle 2, Assistant 1 has particle 1, Assistant 2 has particle 4, and the Receiver has particle 3. As a result, the global quantum state of the whole system is

$$|Q\rangle_{m1234} = |\xi\rangle_m \otimes |\chi^{00}\rangle_{1234}. \quad (4)$$

Eq.(4) can be rewritten into

$$\begin{aligned}
|Q\rangle_{m1234} = \frac{1}{4}|0\rangle_1 &\left\{|\phi^+\rangle_{m2}\left[|0\rangle_4(a|0\rangle_3+b|1\rangle_3)+|1\rangle_4(a|1\rangle_3-b|0\rangle_3)\right]\right.\\
&+|\phi^-\rangle_{m2}\left[|0\rangle_4(a|0\rangle_3-b|1\rangle_3)+|1\rangle_4(a|1\rangle_3+b|0\rangle_3)\right]\\
&+|\varphi^+\rangle_{m2}\left[|0\rangle_4(a|1\rangle_3+b|0\rangle_3)+|1\rangle_4(-a|0\rangle_3+b|1\rangle_3)\right]\\
&+|\varphi^-\rangle_{m2}\left[|0\rangle_4(a|1\rangle_3-b|0\rangle_3)+|1\rangle_4(-a|0\rangle_3-b|1\rangle_3)\right]\right\}\\
+\frac{1}{4}|1\rangle_1 &\left\{|\phi^+\rangle_{m2}\left[|0\rangle_4(a|1\rangle_3+b|0\rangle_3)+|1\rangle_4(a|0\rangle_3-b|1\rangle_3)\right]\right.\\
&+|\phi^-\rangle_{m2}\left[|0\rangle_4(a|1\rangle_3-b|0\rangle_3)+|1\rangle_4(a|0\rangle_3+b|1\rangle_3)\right]\\
&+|\varphi^+\rangle_{m2}\left[|0\rangle_4(a|0\rangle_3+b|1\rangle_3)+|1\rangle_4(-a|1\rangle_3+b|0\rangle_3)\right]\\
&+|\varphi^-\rangle_{m2}\left[|0\rangle_4(a|0\rangle_3-b|1\rangle_3)+|1\rangle_4(-a|1\rangle_3-b|0\rangle_3)\right]\right\}, \quad (5)
\end{aligned}$$

where $|\phi^{\pm}\rangle = \frac{1}{\sqrt{2}}(|00\rangle\pm|11\rangle)$ and $|\varphi^{\pm}\rangle = \frac{1}{\sqrt{2}}(|01\rangle\pm|10\rangle)$.

We put forward a novel quantum teleportation scheme to teleport $|\xi\rangle_m$ from the Sender to the Receiver:

Step 1: Assistant 1 measures particle 1 with the $Z$ basis and sends her measurement result to the Receiver. According to the measurement result of particle 1, the Receiver can know the collapsed state of particles (m,2,3,4).

Step 2: The Sender performs the Bell basis measurement (i.e., $\{|\phi^{\pm}\rangle,|\varphi^{\pm}\rangle\}$) on the particles (m,2) and sends her measurement result to the Receiver. According to the Bell basis measurement result of particles (m,2), the Receiver can know the collapsed state of particles (3,4).

Step 3: Assistant 2 measures particle 4 with the $Z$ basis and sends her measurement result to the Receiver. The Receiver can derive the collapsed state of particle 3 from the measurement result of particle 4. According to Table 1, the Receiver can perform the corresponding unitary operation on particle 3 to recover $|\xi\rangle_m$ in each case. Note that in Table 1, it has $I=|0\rangle\langle 0|+|1\rangle\langle 1|$, $\delta_x=|0\rangle\langle 1|+|1\rangle\langle 0|$, $i\delta_y=|0\rangle\langle 1|-|1\rangle\langle 0|$ and $\delta_z=|0\rangle\langle 0|-|1\rangle\langle 1|$.

**Table 1** The corresponding relations among different parameters in the proposed quantum teleportation scheme

| Assistant 1's Z basis measurement result on particle 1 | The Sender's Bell basis measurement result on particles (m,2) | Assistant 2's Z basis measurement result on particle 4 | The collapsed state of particle 3 | The Receiver's unitary operation |
|---|---|---|---|---|
| $|0\rangle$ | $|\phi^+\rangle$ | $|0\rangle$ | $a|0\rangle+b|1\rangle$ | $I$ |
| | | $|1\rangle$ | $a|1\rangle-b|0\rangle$ | $i\delta_y$ |
| | $|\phi^-\rangle$ | $|0\rangle$ | $a|0\rangle-b|1\rangle$ | $\delta_z$ |
| | | $|1\rangle$ | $a|1\rangle+b|0\rangle$ | $\delta_x$ |
| | $|\varphi^+\rangle$ | $|0\rangle$ | $a|1\rangle+b|0\rangle$ | $\delta_x$ |
| | | $|1\rangle$ | $-a|0\rangle+b|1\rangle$ | $\delta_z$ |
| | $|\varphi^-\rangle$ | $|0\rangle$ | $a|1\rangle-b|0\rangle$ | $i\delta_y$ |
| | | $|1\rangle$ | $-a|0\rangle-b|1\rangle$ | $I$ |
| $|1\rangle$ | $|\phi^+\rangle$ | $|0\rangle$ | $a|1\rangle+b|0\rangle$ | $\delta_x$ |
| | | $|1\rangle$ | $a|0\rangle-b|1\rangle$ | $\delta_z$ |
| | $|\phi^-\rangle$ | $|0\rangle$ | $a|1\rangle-b|0\rangle$ | $i\delta_y$ |
| | | $|1\rangle$ | $a|0\rangle+b|1\rangle$ | $I$ |
| | $|\varphi^+\rangle$ | $|0\rangle$ | $a|0\rangle+b|1\rangle$ | $I$ |
| | | $|1\rangle$ | $-a|1\rangle+b|0\rangle$ | $i\delta_y$ |
| | $|\varphi^-\rangle$ | $|0\rangle$ | $a|0\rangle-b|1\rangle$ | $\delta_z$ |
| | | $|1\rangle$ | $-a|1\rangle-b|0\rangle$ | $\delta_x$ |

## 2.2 A novel semiquantum proxy blind signature protocol with quantum teleportation based on $\chi$ states

The proposed semiquantum proxy blind signature protocol has five participants. Alice is the original message owner who owns the original message need be signed; Bob is the original signer and want to delegate the signing authority to a proxy signer; the proxy signer, David, is authorized to perform signature instead of the original signer; the signature verifier, Charlie, helps verify the validity of the signature; Trent is a third party who helps accomplish the signing process. Alice, Trent and David are quantum participants with full quantum abilities, while Bob and Charlie are semiquantum participants with limited quantum capabilities.

The proposed semiquantum proxy blind signature protocol requires Bob to authorize David to sign the signature on the basis that the content of Alice's original message is blind to all of Bob, Charlie, David and Trent.

The proposed semiquantum proxy blind signature protocol consists of four stages: the initializing phase, the blindness phase, the authorization and signing phase, and the verifying phase. Note that throughout this protocol, $|0\rangle$ and $|1\rangle$ correspond to the classical bits 0 and 1, respectively; and $|\phi^+\rangle, |\phi^-\rangle, |\varphi^+\rangle$ and $|\varphi^-\rangle$ correspond to the classical bits 00, 01, 10 and 11, respectively. For clarity, the information transmissions among different participants of this protocol are shown in Fig.1.

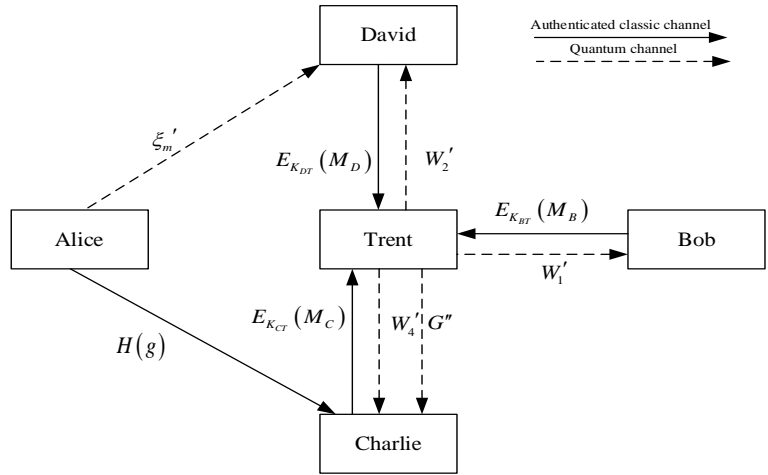

Fig.1 The information transmissions among different participants

### 2.2.1 Initializing phase

**Step 1**: Trent shares a private key $K_{DT}$ of length $2n$ with David via the BB84 QKD scheme [2], where $K_{DT} = \{K_{DT_1}, K_{DT_2}, \ldots, K_{DT_{2n}}\}$, $K_{DT_j} \in \{0,1\}$, $j = 1, 2, \ldots, 2n$. Trent shares a private key $K_{BT}$ of length $n$ with Bob via a secure SQKD scheme [19], where $K_{BT} = \{K_{BT_1}, K_{BT_2}, \ldots, K_{BT_n}\}$, $K_{BT_i} \in \{0,1\}$, $i = 1, 2, \ldots, n$. Trent shares a private key $K_{CT}$ of length $n$ with Charlie via a secure SQKD scheme [19], where

$K_{CT} = \{K_{CT_1}, K_{CT_2}, \ldots, K_{CT_n}\}$, $K_{CT_i} \in \{0,1\}$, $i = 1, 2, \ldots, n$. Alice generates a private key $K_A$ of length $n$ with a random number generator, where $K_A = \{K_{A_1}, K_{A_2}, \ldots, K_{A_n}\}$, $K_{A_i} \in \{0,1\}$, $i = 1, 2, \ldots, n$.

**Step 2**: Trent prepares $n$ quantum states all in the state of $|\chi^{00}\rangle_{1234}$. Let $W_1, W_2, W_3$ and $W_4$ represent the particle sequences composed by all particle 1, particle 2, particle 3 and particle 4 of these $n$ quantum states, respectively. Trent prepares enough decoy particles, each of which is randomly in one of the four states $\{|0\rangle, |1\rangle, |+\rangle, |-\rangle\}$, where $|\pm\rangle = \frac{1}{\sqrt{2}}(|0\rangle \pm |1\rangle)$, and then randomly inserts them into $W_1$ to get $W_1'$. Trent prepares enough decoy particles, each of which is also randomly in one of the four states $\{|0\rangle, |1\rangle, |+\rangle, |-\rangle\}$, and then randomly inserts them into $W_4$ to get $W_4'$. Trent prepares enough decoy particles, each of which is also randomly in one of the four states $\{|0\rangle, |1\rangle, |+\rangle, |-\rangle\}$, and randomly inserts them into $W_2$ to get $W_2'$. Finally, Trent keeps $W_3$ to himself and sends $W_1'$ to Bob, $W_2'$ to David and $W_4'$ to Charlie.

**Step 3**: Alice's classical message need be signed is represented as $g_A = \{g_{A_1}, g_{A_2}, \ldots, g_{A_n}\}$, where $g_{A_i} \in \{0,1\}$, $i = 1, 2, \ldots, n$. Alice computes $g_i = g_{A_i} \oplus K_{A_i}$ to obtain the blind classical message $g$, where $g = \{g_1, g_2, \ldots, g_n\}$, $g_i \in \{0,1\}$, $i = 1, 2, \ldots, n$. $H(\cdot)$ is a collision-resistant one-way hash function shared only between Alice and Charlie. Alice computes $H(g)$ and sends it to Charlie.

### 2.2.2 Blindness phase

Alice turns $g$ into the quantum state sequence $G = \{|g_1\rangle, |g_2\rangle, \ldots, |g_n\rangle\}$, and prepares the single-particle state sequence $\xi_m = \{|\xi_1\rangle_m, |\xi_2\rangle_m, \ldots, |\xi_n\rangle_m\}$ according to the rule defined in Table 2. Concretely speaking, when $|g_i\rangle = |0\rangle$, Alice prepares $|\xi_i\rangle_m = \frac{1}{\sqrt{2}}(|0\rangle_m + |1\rangle_m)$; and when $|g_i\rangle = |1\rangle$, Alice prepares $|\xi_i\rangle_m = \frac{1}{\sqrt{2}}(|0\rangle_m - |1\rangle_m)$.

Table 2  The rule of preparing $|\xi_i\rangle_m$ for Alice

| $|g_i\rangle$ | $|\xi_i\rangle_m$ |
|---|---|
| $|0\rangle$ | $\frac{1}{\sqrt{2}}(|0\rangle_m + |1\rangle_m)$ |
| $|1\rangle$ | $\frac{1}{\sqrt{2}}(|0\rangle_m - |1\rangle_m)$ |

### 2.2.3 Authorization and signing phase

**Step 1**: Alice prepares enough decoy particles, each of which is randomly in one of the four states $\{|0\rangle, |1\rangle, |+\rangle, |-\rangle\}$, and randomly inserts them into $\xi_m$ to get $\xi_m'$. Then,

Alice sends $\xi'_m$ to David.

**Step 2**: After David receives $\xi'_m$ from Alice, Alice announces the positions and the measuring basis of decoy particles. David measures the decoy particles with the corresponding measuring basis and informs Alice of his measurement results. Alice calculates the error rate by comparing David's measurement results on decoy particles with their corresponding original states. If the error rate is low enough, David will ask Bob for approval to generate signature; otherwise, the protocol will be terminated.

**Step 3**: After Bob receives $W'_1$ from Trent, Trent announces the positions of decoy particles. For each decoy particle, Bob randomly chooses either to measure it with the $Z$ basis (referred as SIFT) or to reflect it back to Trent (referred as CTRL). Bob randomly reorders the reflected particles. Trent collects the reflected particles from Bob temporarily and announces which particles were prepared by him in the $Z$ basis.

Bob announces which particles were performed by him with the CTRL operations and in which order they were reflected. Trent then measures all returned particles with their preparing basis. Trent calculates the error rate on the reflected particles. If it is low enough, the protocol will be continued; otherwise, the protocol will be terminated.

Bob publishes his $Z$ basis measurement results. Trent checks the error rate on the particles which were both prepared by him in the $Z$ basis and measured by Bob with the $Z$ basis. If it is low enough, the protocol will be continued; otherwise, the protocol will be terminated.

**Step 4**: After performing the eavesdropping detection, Bob recovers $W_1$. Bob receives David's request and agrees to authorize the signature. Then Bob measures the particles of $W_1$ with the $Z$ basis, and obtains the classical bits corresponding to his measurement results, i.e., $M_B = \{M_B(1), M_B(2), ..., M_B(n)\}$, where $M_B(i) \in \{0,1\}$, $i = 1, 2, ..., n$. Bob encrypts $M_B$ with $K_{BT}$ as $E_{K_{BT}}[M_B]$ and sends $E_{K_{BT}}[M_B]$ to Trent.

**Step 5**: Trent decrypts $E_{K_{BT}}[M_B]$ from Bob to get $M_B$ and informs David to sign.

**Step 6**: David receives $W'_2$ from Trent and recovers $W_2$ after performing the same eavesdropping detection as that in Step 2. David gets $\xi_m$ after the eavesdropping detection in Step 2. After receiving the signature notification from Trent, David performs the Bell basis measurement on the particle pair of $|\xi_i\rangle_m$ and the $i$th particle of $W_2$, and obtains the classical bits corresponding to his measurement result, i.e., $M_D(i)$, where $M_D(i) \in \{00, 01, 10, 11\}$, $i = 1, 2, ..., n$. David encrypts $M_D = \{M_D(1), M_D(2), ..., M_D(n)\}$ with $K_{DT}$ as $E_{K_{DT}}(M_D)$ and then sends $E_{K_{DT}}(M_D)$ to Trent.

**Step 7**: Trent decrypts $E_{K_{DT}}(M_D)$ from David to get $M_D$. Trent tells Charlie to measure the particles of $W_4$.

**Step 8**: Charlie receives $W_4'$ from Trent and recovers $W_4$ after performing the same eavesdropping detection as that in Step 3. Then Charlie measures the particles of $W_4$ with the $Z$ basis, and obtains the classical bits corresponding to her measurement results, i.e., $M_C = \{M_C(1), M_C(2), ..., M_C(n)\}$, where $M_C(i) \in \{0,1\}$, $i = 1, 2, ..., n$. Charlie encrypts $M_C$ with $K_{CT}$ as $E_{K_{CT}}(M_C)$ and then sends $E_{K_{CT}}(M_C)$ to Trent.

**Step 9**: Trent decrypts $E_{K_{CT}}(M_C)$ from Charlie to obtain $M_C$. Then, according to Table 1, Trent performs the corresponding unitary operation to recover $\xi_m$. Afterward, Trent prepares $G'$ according to $\xi_m$, where $G' = \{|g_1'\rangle, |g_2'\rangle, ..., |g_n'\rangle\}$, $g_i' \in \{0,1\}$, $i = 1, 2, ..., n$. Concretely, when $|\xi_i\rangle_m = \frac{1}{\sqrt{2}}(|0\rangle_m + |1\rangle_m)$, Trent prepares $|g_i'\rangle = |0\rangle$; and when $|\xi_i\rangle_m = \frac{1}{\sqrt{2}}(|0\rangle_m - |1\rangle_m)$, Trent prepares $|g_i'\rangle = |1\rangle$. Afterward, Trent prepares enough decoy particles, each of which is randomly in one of the four states $\{|0\rangle, |1\rangle, |+\rangle, |-\rangle\}$, and inserts them into $G'$ to get $G''$. At last, Trent sends $G''$ to Charlie.

#### 2.2.4 Verifying phase

**Step 1**: Charlie receives $G''$ from Trent and recovers $G'$ after performing the same eavesdropping detection as that in Step 3 of Sect.2.2.3. Charlie measures the particles of $G'$ with the $Z$ basis and obtains $g'$ according to her $Z$ basis measurement results, where $g' = \{g_1', g_2', ..., g_n'\}$, $g_i' \in \{0,1\}$, $i = 1, 2, ..., n$.

**Step 2**: Charlie judges whether $H(g')$ and $H(g)$ are identical or not. If they are identical, Charlie declares that the signature is valid; otherwise, she announces that the signature is invalid.

## 3 Discussion

### 3.1 Blindness

In our protocol, Bob measures the particles of $W_1$ with the $Z$ basis, and obtains $M_B$. However, it is still helpless for Bob to obtain $g_A$.

David receives $\xi_m$ and further $g$ from Alice. Alice's original message $g_A$ is encrypted by $K_A$ to produce the blind message $g$. $K_A$ is only known by Alice. As a result, David cannot obtain $K_A$, which means that $g_A$ is blind to David.

During the transmission process of $\xi_m'$ from Alice to David, our protocol employs the decoy particle technology [26,27], which can effectively resist the eavesdropping behavior from an eavesdropper. As a result, an eavesdropper cannot get $\xi_m$ and further $g$ without being discovered. Moreover, during the transmission process of $H(g)$ from Alice to Charlie, due to the one-way property of the collision-resistant hash function $H(\cdot)$, even if an eavesdropper intercepts $H(g)$, she still cannot restore $g$, let alone $g_A$.

Thus, $g_A$ is also blind to an eavesdropper.

Trent recovers $\xi_m$ and prepares $G'$ according to it in Step 9. Even though $G' = G$, due to having no knowledge about $K_A$, Trent still cannot obtain $g_A$.

Charlie measures the particles of $G'$ with the Z basis and obtains $g'$ according to her Z basis measurement results. Even though $g' = g$, due to no access to $K_A$, Charlie still has no knowledge about $g_A$.

To sum up, our protocol has complete blindness.

### 3.2 Unforgeability

In the signing phase, because of the unconditional security of QKD, $K_{DT}$ cannot be obtained by any dishonest inside attacker except Trent and David. Hence, an inside attacker except Trent and David cannot decrypt out $M_D$, even though she intercepts $E_{K_{DT}}(M_D)$ from David to Trent. Therefore, an inside attacker except Trent and David cannot forge the signature.

In addition, if an outside attacker tries to forge the signature, she will need to know $g$ or $M_D$. However, $g$ is blind to the outside attacker, according to Sect.3.2. Moreover, since an outside attacker cannot know $K_{DT}$, even though she intercepts $E_{K_{DT}}(M_D)$ from David to Trent, she still cannot decrypt out $M_D$. Therefore, an outside attacker also cannot forge the signature.

To sum up, our protocol is unforgeable.

### 3.3 Non-repudiation

The repudiation of our protocol includes the original signer repudiating authorization to the proxy signer, the proxy signer repudiating the generation of signature and the third party repudiating the acceptance of signature.

Firstly, the original signer, Bob, cannot repudiate authorizing the proxy signer, David. The recovery of $\xi_m$ by Trent requires him to perform the corresponding unitary operations according to Table 1. Bob's measurement results on the particles of $W_1$ should be known by Trent before performing the corresponding unitary operation. After Trent know Bob's measurement results on the particles of $W_1$, he informs David to sign. Therefore, when Charlie successfully verifies the validity of signature after Trent recovers $\xi_m$, Bob cannot repudiate authorizing David to sign.

Secondly, the proxy signer, David, cannot deny that the signature was generated by him. The recovery of $\xi_m$ by Trent requires him to perform the corresponding unitary operations according to Table 1. It is necessary for him to know David's Bell basis measurement result on the particle pair of $|\xi_i\rangle_m$ and the $i$ th particle of $W_2$ before performing the corresponding unitary operation. Therefore, when Charlie successfully verifies the validity of signature after Trent recovers $\xi_m$, David cannot repudiate generating the signature by measuring the particle pair of $|\xi_i\rangle_m$ and the $i$th particle of $W_2$.

Thirdly, the third party, Trent, cannot repudiate the acceptance of signature. In order to verify the validity of signature, Charlie needs to obtain $g'$ and further require Trent to recover $\xi_m$. In order to recover $\xi_m$, Trent needs to accept David's Bell basis measurement result on the particle pair of $|\xi_i\rangle_m$ and the $i$th particle of $W_2$. Therefore, when Charlie successfully verifies the validity of signature, Trent cannot repudiate accepting David's signature.

### 3.4 The entangle-measure attack

With respect to the qubit transmissions of our protocol, Trent sends $W_1'$ to Bob, $W_2'$ to David and $W_4'$ and $G''$ to Charlie, while Alice sends $\xi_m'$ to David. Apparently, both the eavesdropping detection between Trent and Bob for the transmission of $W_1'$ and the eavesdropping detection between Trent and Charlie for the transmissions of $W_4'$ and $G''$ are same as that of randomization-based SQKD protocol in Ref.[20], which has been proven to have complete robustness. Hence, we only need to analyze the security for transmissions of $W_2'$ from Trent to David and $\xi_m'$ from Alice to David. Since both the transmission of $W_2'$ from Trent to David and the transmission of $\xi_m'$ from Alice to David employ the same decoy particle technology, in the following, without loss of generality, we only focus on analyzing the security for transmission of $\xi_m'$ from Alice to David.

During the transmission of $\xi_m'$ from Alice to David, Eve may try to get some useful information through her entangle-measure attack. That is, Eve may entangle her auxiliary particle $|\varepsilon\rangle_E$ with the particle of $\xi_m'$ sent out from Alice by imposing $\hat{E}$. The effect of $\hat{E}$ on the qubits $|0\rangle$ and $|1\rangle$ can be expressed as:

$$\hat{E}(|0\rangle|\varepsilon\rangle_E) = \alpha_{00}|0\rangle|\varepsilon_{00}\rangle + \alpha_{01}|1\rangle|\varepsilon_{01}\rangle, \qquad (6)$$

$$\hat{E}(|1\rangle|\varepsilon\rangle_E) = \alpha_{10}|0\rangle|\varepsilon_{10}\rangle + \alpha_{11}|1\rangle|\varepsilon_{11}\rangle, \qquad (7)$$

where $|\varepsilon_{00}\rangle, |\varepsilon_{01}\rangle, |\varepsilon_{10}\rangle, |\varepsilon_{11}\rangle$ are Eve's probe state determined by $\hat{E}$, $|\alpha_{00}|^2 + |\alpha_{01}|^2 = 1$ and $|\alpha_{10}|^2 + |\alpha_{11}|^2 = 1$.

According to Eq.(6) and Eq.(7), the effect of $\hat{E}$ on the qubits $|+\rangle$ and $|-\rangle$ can be expressed as:

$$\hat{E}|+\rangle|\varepsilon\rangle = \frac{1}{\sqrt{2}}(\alpha_{00}|0\rangle|\varepsilon_{00}\rangle + \alpha_{01}|1\rangle|\varepsilon_{01}\rangle + \alpha_{10}|0\rangle|\varepsilon_{10}\rangle + \alpha_{11}|1\rangle|\varepsilon_{11}\rangle)$$

$$= \frac{1}{2}|+\rangle(\alpha_{00}|\varepsilon_{00}\rangle + \alpha_{01}|\varepsilon_{01}\rangle + \alpha_{10}|\varepsilon_{10}\rangle + \alpha_{11}|\varepsilon_{11}\rangle)$$

$$+\frac{1}{2}|-\rangle\left(\alpha_{00}|\varepsilon_{00}\rangle-\alpha_{01}|\varepsilon_{01}\rangle+\alpha_{10}|\varepsilon_{10}\rangle-\alpha_{11}|\varepsilon_{11}\rangle\right), \tag{8}$$

$$\hat{E}(|-\rangle|\varepsilon\rangle_E) = \frac{1}{\sqrt{2}}\left(\alpha_{00}|0\rangle|\varepsilon_{00}\rangle+\alpha_{01}|1\rangle|\varepsilon_{01}\rangle-\alpha_{10}|0\rangle|\varepsilon_{10}\rangle-\alpha_{11}|1\rangle|\varepsilon_{11}\rangle\right)$$

$$=\frac{1}{2}|+\rangle\left(\alpha_{00}|\varepsilon_{00}\rangle+\alpha_{01}|\varepsilon_{01}\rangle-\alpha_{10}|\varepsilon_{10}\rangle-\alpha_{11}|\varepsilon_{11}\rangle\right)$$

$$+\frac{1}{2}|-\rangle\left(\alpha_{00}|\varepsilon_{00}\rangle-\alpha_{01}|\varepsilon_{01}\rangle-\alpha_{10}|\varepsilon_{10}\rangle+\alpha_{11}|\varepsilon_{11}\rangle\right). \tag{9}$$

In order for Eve not being detected by Alice and David, Eq.(6) and Eq.(7) should satisfy that

$$\alpha_{01}=\alpha_{10}=0; \tag{10}$$

in the meanwhile, Eq.(8) and Eq.(9) should satisfy that

$$\alpha_{00}|\varepsilon_{00}\rangle-\alpha_{01}|\varepsilon_{01}\rangle+\alpha_{10}|\varepsilon_{10}\rangle-\alpha_{11}|\varepsilon_{11}\rangle=0, \tag{11}$$

$$\alpha_{00}|\varepsilon_{00}\rangle+\alpha_{01}|\varepsilon_{01}\rangle-\alpha_{10}|\varepsilon_{10}\rangle-\alpha_{11}|\varepsilon_{11}\rangle=0. \tag{12}$$

Inserting Eq.(10) into Eq.(11) and Eq.(12) produces

$$\alpha_{00}|\varepsilon_{00}\rangle=\alpha_{11}|\varepsilon_{11}\rangle=|\tau\rangle. \tag{13}$$

Inserting Eq.(10) and Eq.(13) into Eq.(6) and Eq.(7) produces

$$\hat{E}(|0\rangle|\varepsilon\rangle_E)=|0\rangle|\tau\rangle, \tag{14}$$

$$\hat{E}(|1\rangle|\varepsilon\rangle_E)=|1\rangle|\tau\rangle. \tag{15}$$

Inserting Eq.(10) and Eq.(13) into Eq.(8) and Eq.(9) produces

$$\hat{E}(|+\rangle|\varepsilon\rangle_E)=|+\rangle|\tau\rangle, \tag{16}$$

$$\hat{E}(|-\rangle|\varepsilon\rangle_E)=|-\rangle|\tau\rangle. \tag{17}$$

It can be concluded from Eqs.(14-17) that in order not to be detected, Eve's auxiliary particle should always be in the state of $|\tau\rangle$, which means that Eve gets nothing useful by measuring her launching auxiliary particle with the correct measuring basis.

## 4  Comparison

The qubit efficiency of the proposed semiquantum proxy blind signature protocol can be calculated as [28]

$$\eta=\frac{q_s}{q_t+q_c}, \tag{18}$$

where $q_s$, $q_t$ and $q_c$ represent the signature length, the number of qubits consumed and the number of classical bits used during classical communication processes, respectively. The qubits and the classical bits consumed during eavesdropping detection processes are ignored in the following.

In our protocol, after Alice authorizes David to sign the signature, David generates $M_D$ by performing the Bell basis measurement on the particle pair of $|\xi_i\rangle_m$ and the $i$ th particle of $W_2$, where $i = 1, 2, \ldots, n$, thus it has $q_s = 2n$. Trent prepares $n$ quantum states all in the state of $|\chi^{00}\rangle_{1234}$; Alice prepares $\xi_m$ according to $G$; Trent prepares $G'$ according to $\xi_m$; Trent shares $K_{DT}$ with David via the BB84 QKD scheme; moreover, Trent shares $K_{BT}$ and $K_{CT}$ with Bob and Charlie via the SQKD scheme in Ref.[19], respectively; thus it has $q_t = 4n + n + n + 8n + 8n + 8n = 30n$. Alice sends $H(g)$ to Charlie; moreover, Bob, David and Charlie sends $E_{K_{BT}}[M_B]$, $E_{K_{DT}}(M_D)$ and $E_{K_{CT}}(M_C)$ to Trent, respectively, hence it has $q_c = l + n + 2n + n = l + 4n$, where $l$ is the length of output bit string of $H(\cdot)$. Hence, the qubit efficiency of our protocol is $\eta = \dfrac{2n}{30n + l + 4n} = \dfrac{2n}{34n + l}$.

The comparison results of our protocol and previous semiquantum signature protocols based on multi-particle quantum entangled states are summarized in Table 3.

In the semiquantum signature protocol of Ref.[21], Alice produces her signature $S_A$, while Bob produces his signature $S_B$, thus it has $q_s = 2n$. Alice prepares $n$ W states, sequence $s_m$ and sequence $S'_A$; Bob prepares $n$ W states and sequence $S'_B$; moreover, Alice and Charlie pre-share a semiquantum key $K_{AC}$ via the SQKD scheme in Ref.[19], while Bob and Charlie pre-share a semiquantum key $K_{BC}$ via the SQKD scheme in Ref.[19]; thus it has $q_t = 3n + n + n + 3n + n + 8n + 8n = 25n$. Alice sends $\{S_A, A\}$ and $K_{AC} \oplus M_A$ to Charlie; moreover, Bob sends $\{S_B, B\}$ and $K_{BC} \oplus M_B$ to Charlie; hence it has $q_c = 2n + n + 2n + n = 6n$. Hence, the qubit efficiency of the semiquantum signature protocol in Ref.[21] is $\eta = \dfrac{2n}{25n + 6n} = \dfrac{2}{31}$.

In the semiquantum signature protocol of Ref.[23], Bob produces his signature $S_B = E_{K_{AB}}[|B\rangle]$, thus it has $q_s = n$. Alice prepares $n$ five-particle GHZ states and quantum state sequence $|M\rangle$; Bob prepares quantum state sequence $|B\rangle$; Alice shares $K_{AB}$ with Bob via the BB84 QKD scheme; Alice shares $K_{AC}$ with Charlie via the SQKD scheme in Ref.[19], while Bob shares $K_{BC}$ with Charlie via the SQKD scheme in Ref.[19]; thus it has $q_t = 5n + n + n + 4n + 8n + 8n = 27n$. Charlie announces $k_{BC} \oplus R_c$ to Bob; Bob computes $R_c = k_{BC} \oplus (k_{BC} \oplus R_c)$, compares $R_b$ and $R_c$, and then announces Charlie the positions where the comparison results are same and opposite between $R_b$ and $R_c$; hence it has $q_c = n + n = 2n$. Hence, the qubit efficiency of the semiquantum signature protocol in Ref.[23] is $\eta = \frac{n}{27n+2n} = \frac{1}{29}$.

**Table 3** Comparison results of our protocol and previous semiquantum signature protocols based on multi-particle quantum entangled states

|  | The protocol of Ref.[21] | The protocol of Ref.[23] | Our protocol |
| --- | --- | --- | --- |
| Quantum resource | W states and single-particle states | Five-particle GHZ states and single-particle states | $\chi$ states and single-particle states |
| Semiquantum parties | The signature verifier | The signature verifier | The original signer and the signature verifier |
| The number of original message owner | Two | One | One |
| The number of proxy signer | Zero | Zero | One |
| Eavesdropping check | No | Yes | Yes |
| Quantum measurements for quantum parties | Three-particle quantum entangled state measurements and Z basis measurements | Z basis measurements | Bell basis measurements and Z basis measurements |
| Quantum measurements for semiquantum parties | Z basis measurements | Z basis measurements | Z basis measurements |
| Usage of pre-shared QKD keys or SQKD keys | Yes | Yes | Yes |
| Usage of quantum teleportation | Yes | No | Yes |
| Usage of unitary operations | Yes | Yes | Yes |
| Qubit efficiency | $\frac{2}{31}$ | $\frac{1}{29}$ | $\frac{2n}{34n+l}$ |

According to Table 3, as for quantum resource, our protocol exceeds the protocol of Ref.[23], as four-particle quantum entangled states are much easier to prepare than five-particle quantum entangled states; since there is no proxy signer in the protocols of Ref.[21] and Ref.[23], these two protocols belong to semiquantum blind signature protocol, but our protocol is a semiquantum proxy blind signature protocol with one proxy signer; as there is no eavesdropping check process for qubit transmissions in the protocol of Ref.[21], it has great security loopholes, but our protocol doesn't have this problem, because of employing the eavesdropping check processes for qubit transmissions; with respect to quantum measurements for quantum parties, our protocol exceeds the protocol of Ref.[21], as Bell basis measurements are much easier to realize than three-particle quantum entangled state measurements; and as long as $l < 24n$, our protocol has a higher qubit efficiency than the protocol of Ref.[23].

## 5 Conclusion

In this paper, a novel semiquantum proxy blind signature protocol with quantum teleportation based on $\chi$ states is designed. In order to make her original classical message blind, the original message owner, Alice, converts it into a corresponding single-particle quantum superposition state sequence. Based on quantum teleportation with $\chi$ states and single-particle quantum superposition states, the original signer, Bob, authorizes the proxy signer, David, through $Z$ basis measurements; David signs the signature through Bell basis measurements; the third party, Trent, restores Alice's single-particle superposition state sequence; and the signature verifier, Charlie, verifies the signature after receiving a new single-particle state sequence corresponding to Alice's single-particle superposition state sequence from Trent. It is validated in detail that our protocol satisfies complete blindness, unforgeability, non-repudiation and security. Because Bob and Charlie are semiquantum participants with limited quantum capabilities, our protocol is likely to need less quantum resources and be easier to implement in reality, compared with many previous quantum proxy blind signature protocols.

### Acknowledgments

Funding by the National Natural Science Foundation of China (Grant No.62071430 and No.61871347) and the Fundamental Research Funds for the Provincial Universities of Zhejiang (Grant No.JRK21002) is gratefully acknowledged.